\documentclass[preprint]{iucr}              

\usepackage{graphicx}
\usepackage{xspace}
\usepackage{threeparttable}
\usepackage{geometry}
\usepackage{xcolor}
\usepackage{enumitem}

\usepackage{amsmath}
\usepackage{booktabs}
\usepackage{multirow}
\usepackage[normalem]{ulem}
\usepackage{tikz}

\graphicspath{{./figures/}}

\newcommand{\be}{\begin{enumerate}[wide, labelwidth=!, labelindent=0pt,
		label=\textbf{\textcolor{blue}{\arabic*}.}]}

\newcommand{\bei}{\begin{enumerate}}
\newcommand{\ee}{\end{enumerate}}
\newcounter{saveenumi}

\let\vec\mathbf

\newcommand{\dd}{\mathrm{d}}

\newcommand{\fig}[1]{Fig.~\ref{fig:#1}}

\newcommand{\nba}[1]{}

\newcommand{\ppdf}{pwPDF\xspace}
\newcommand{\tpdf}{txPDF\xspace}

\def\href#1{\relax}\let\foo\caption
\ifPDF
\RequirePackage{hyperref}
\PassOptionsToPackage{pdftex,bookmarksopen,bookmarksnumbered}{hyperref}
\fi
\let\caption\foo
\paperprodcode{a000000}      
\paperref{xx9999}            
\papertype{IU}               
\paperlang{english}          
\journalcode{A}             
\journalyr{2003}
\journalreceived{\relax}
\journalaccepted{\relax}
\journalonline{\relax}

\begin{document}                  
	
	
	
	\title{Real space texture and pole figure analysis using the three-dimensional pair distribution function on a platinum thin film}
	\shorttitle{Textured 3D PDF}
	
	\author[a, b]{Sani~Y.}{Harouna-Mayer}
	\author[c]{Songsheng}{Tao}
	\author[c]{ZiZhou}{Gong}
	\author[d]{Martin}{v. Zimmermann}
	\author[a, b]{Dorota}{Koziej}
	\author[d]{Ann-Christin}{Dippel*}
	\cauthor[c, e]{Simon~J.~L.}{Billinge*}{sb2896@columbia.edu}
	
	\aff[a]{Institute for Nanostructure and Solid-State Physics, Center for Hybrid Nanostructures (CHyN), University of Hamburg, \city{Hamburg}, \country{Germany}}
	\aff[b]{The Hamburg Center for Ultrafast Imaging, \city{Hamburg}, \country{Germany}}
	\aff[c]{Department of Applied Physics and Applied Mathematics, Columbia University, \city{New York}, \country{USA}}
	\aff[d]{Deutsches Elektronen-Synchrotron DESY, \city{Hamburg}, \country{Germany}}
	\aff[e]{Condensed Matter Physics and Materials Science Department, Brookhaven National Laboratory, \city{Upton}, NY 11973 \country{USA}}
	
	\keyword{texture}
	\keyword{3D PDF}
	\keyword{real space pole figure}
	\keyword{real space fiber plot}
	
	\maketitle                        
	
	
\begin{abstract}
An approach is described for studying texture in nanostructured materials.  The approach implements the real space texture PDF, \tpdf, laid out in [Gong and Billinge (2018) arXiv:1805.10342 [cond-mat]].  It is demonstrated on a fiber textured polycrystalline Pt thin film.
The approach uses 3D PDF methods to reconstruct the orientation distribution function (ODF) of the powder crystallites from a set of diffraction patterns taken at different tilt angles of the substrate with respect to the incident beam directly from the 3D PDF of the sample.  A real space equivalent of the reciprocal space pole figure is defined in terms of interatomic vectors in the PDF and computed for various interatomic vectors in the Pt film.  Further, it is shown how a valid isotropic PDF may be obtained from a weighted average over the tilt series, and the measurement conditions for the best approximant to the isotropic PDF from a single exposure, which for the case of the fiber textured film was in a nearly grazing incidence orientation of around 10 degrees.  Finally, we describe an open source Python software package, Fourigui, that may be used to help in studies of texture from 3D reciprocal space data, and indeed for Fourier transforming and visualizing 3D PDF data in general.
\end{abstract}
	
	\begin{figure}
		\centering
		\caption{\label{fig:nice_3d_pdf}3D representation of the \tpdf of the textured polycrystalline Pt thin film.}
		\includegraphics[width=0.35\columnwidth]{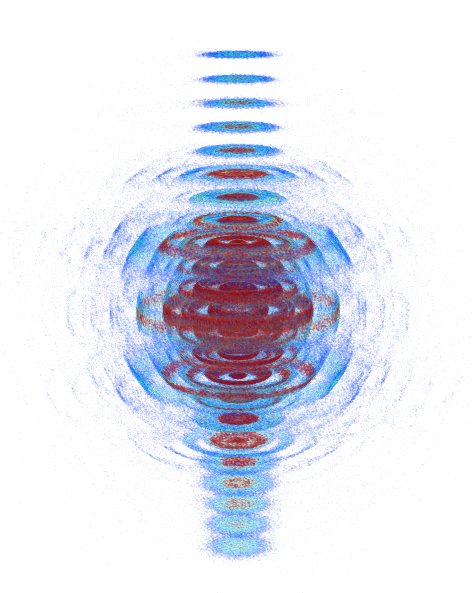}
	\end{figure}
	
\section{Introduction}
	

Atomic pair distribution function (PDF) analysis of x-ray, neutron or electron diffraction data is a powerful method to investigate local structure in materials which can range from amorphous to crystalline~\cite{egami_underneath_2012}.  It is particularly useful for studying structure in small nanoparticles \cite{yang_confirmation_2013, grote_x-ray_2021}, battery materials \cite{shan_framework_2019, hu_oxygen_2021} and thin films \cite{dippel_evolution_2020, song_structural_2021}.
	
	

Conventionally, the PDF assumes the sample to be structurally isotropic, for example, an perfect powder sample.  In general, polycrystalline samples may not always conform to this ideal and have preferred orientation of their grains called a crystallographic texture.   Historically, depending on the extent of the texture, PDF analysis of such samples is not possible or requires tedious sample preparation which results in a loss of all texture information \cite{dippel_local_2019, roelsgaard_time_2019, guo_local_2013}.
	
	

Recently, it was shown how to propagate texture information from samples through the Fourier transform to obtain a 3D PDF of a textured polycrystalline sample, henceforth referred to as \tpdf \cite{gong_atomic_2018}.  This would allow, in principle, the determination of oriental distribution functions that encode the texture directly in real-space, which may be favorable if the powder grains are nanocrystalline.  In the case of weakly textured samples, equations have also been derived to correct the Debye scattering equation for texture information \cite{cervellino_texture_2020}.
	
	
In this paper we explore how to extract texture and structural information in practice, from the \tpdf. After reviewing the mathematical framework of the \tpdf~\cite{gong_atomic_2018}, we describe data acquisition, data processing, texture and structure analysis procedures on the example of a fiber textured platinum thin film.  We show how to obtain real-space pole-figures and fiber plots and how they can be used to get quantitative information about the texture in the sample.  We also explore how conventional untextured PDFs may be recovered from the data for structural modeling.  The experiments are quick, especially at a high power synchrotron, and the analysis is straightforward.  For textured nanostructure, where analysis of well defined Bragg peaks is problematic, this is likely to be a valuable tool for studying texture, and also shows an approach to mitigate texture when the structure of a nanostructured sample is the target.
	
	
\section{Summary of the \tpdf method}
	
	

Here, we summarize the theory of the 3D PDF of textured polycrystalline samples, \tpdf, that was introduced in \cite{gong_atomic_2018}. In contrast to the \tpdf, the conventional 1D PDF is herein referred to as \ppdf to differentiate it.
	

The conventional 1D PDF, \ppdf, from an isotropic powder, $G(r)$, is obtained by a sine Fourier transform of the reduced structure function, $F(Q) = Q[S(Q)-1]$ where the structure function, $S(Q)$, is related to the measured
coherent scattering intensity, $I(Q)$ by \cite{egami_underneath_2012}
	\begin{equation}
		S(Q) = \dfrac{I(Q)}{\langle f(Q) \rangle^{2}},
	\end{equation}
where $Q$ is the magnitude of the scattering vector~\cite{egami_underneath_2012}.  Here, $f(Q)$ is the atomic form factor and the angle brackets indicate an average over all the chemical species present.   The PDF is then obtained via
	\begin{equation}
		\label{eq:1DPDF}
		G(r) = \frac{1}{2\pi}
		\int_{Q_{min}}^{Q_{max}} Q[ S(Q) - 1 ] \sin Qr \: \dd Q.
	\end{equation}
Below we summarize the extension to this equation for the case where the scattering is not isotropic.	

pwPDF measurements are often carried out at synchrotron sources due to the abundance of high energy x-rays. Data is acquired on 2D area detectors. To yield $S(Q)$, the 2D detector image is azimuthally integrated after applying any masks for problematic pixels in the image, resulting in a 1D function of x-ray intensity vs $Q$. The conversion from detector units to $Q$ is done with the help of a calibration measurement using a known material.  Parasitic scattering is removed, such as measured background signals, and other effects such as multiple scattering and incoherent scattering are also removed and the intensity is divided by the average of the atomic form factor for the material resulting in $S(Q)$, which is then propagated through the FT to obtain the \ppdf.  More details may be found in \cite{egami_underneath_2012}.
	
\begin{figure}
		\includegraphics[width=0.6\columnwidth]{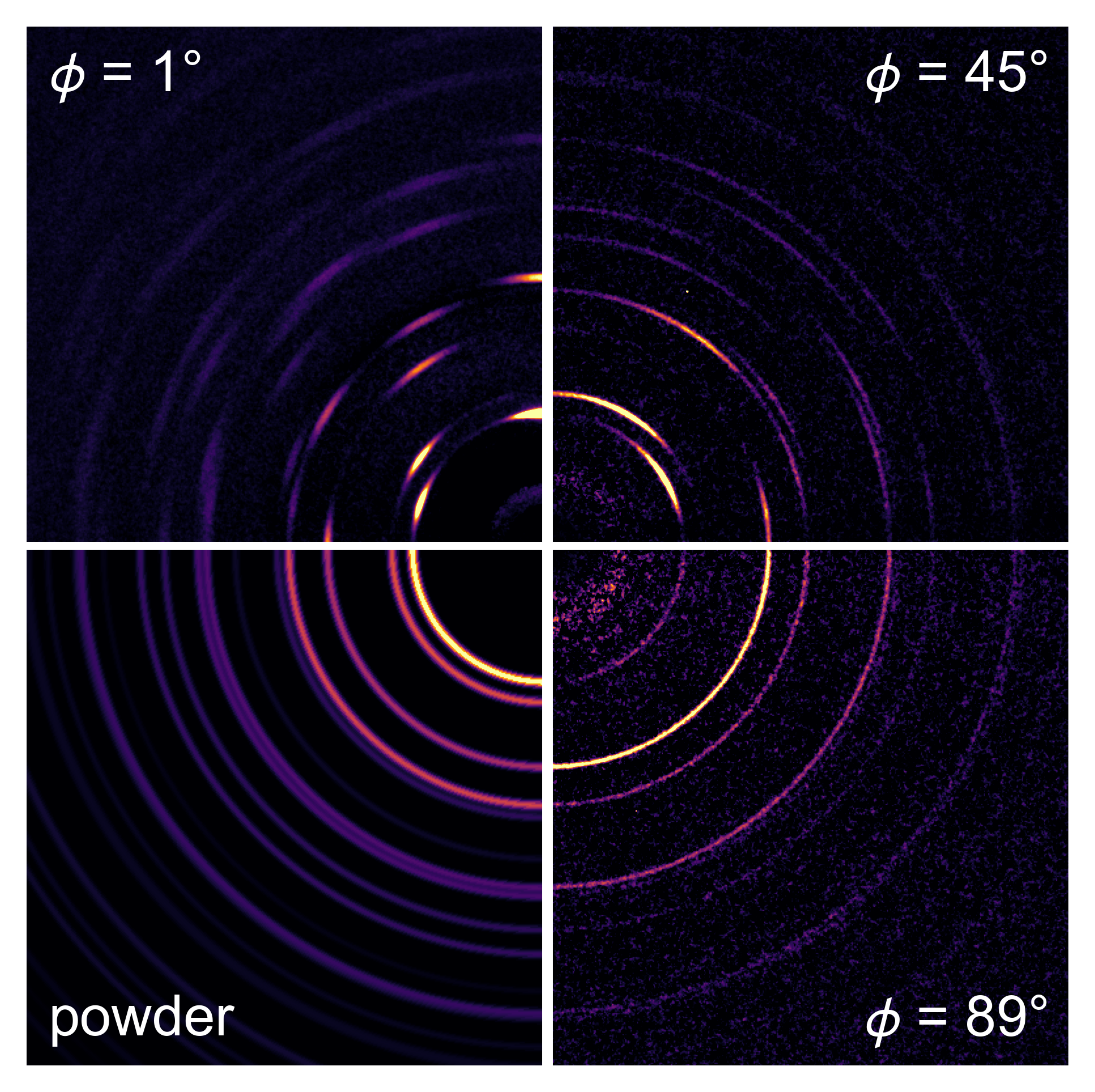}
		\caption{ Sections of detector images of the textured Pt sample at different tilt angles $\mathrm{\phi}$ and a simulated perfect powder Pt detector image. The colorscale is in arbitrary units.}
		\label{fig:pxrd_tex_n_sim}
	\end{figure}

\begin{figure}
		\includegraphics[width=0.6\columnwidth]{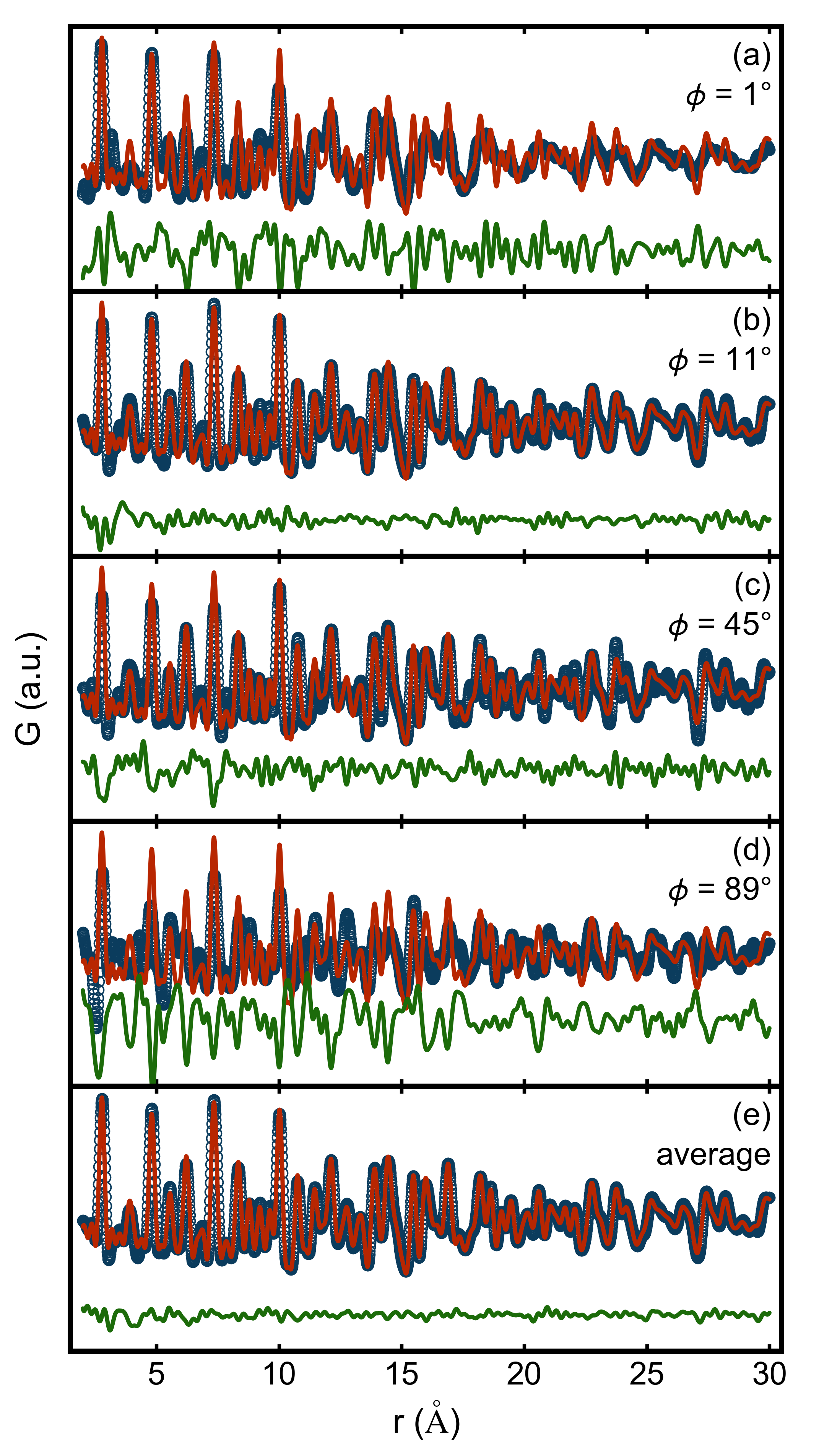}
		\caption{Comparison of measured and best fit calculated PDFs for diffraction images taken at different sample orientations.  The blue circles are the experimental PDFs for sample tilt angles $\mathrm{\phi}$ = 1$^\circ$, 11$^\circ$, 45$^\circ$ and 89$^\circ$ and from the average of the diffraction images from 1$^\circ$ to 89$^\circ$. The red curves are the best fit calculated PDFs from an untextured fcc model. The green curves are the difference between the calculated and measured PDFs. The PDF refinement from the angle-averaged pattern yields a good result with $R\mathrm{_{w}}$ = 12.0\%. Among the single-angle PDFs $\mathrm{\phi}$ = 11$^\circ$ shows the best refinement result ($R\mathrm{_{w}}$ = 18.1\%) and $\mathrm{\phi}$ = 89$^\circ$ shows the worst result ($R\mathrm{_{w}}$ = 63.6\%) (Fig. \ref{fig:pdf_ref_res}).}
		\label{fig:multipanel_pdf}
	\end{figure}

	
In the case of a textured sample, Bragg peak intensities are not uniform in space. The intensities are increased or decreased relative to the average value depending on the orientation of the sample and the direction in space that the intensity is measured.  In the case where the Bragg intensities are measured with a 2D detector, this can result in the diffraction image exhibiting intensity variations around the Debye-Scherrer rings (\fig{pxrd_tex_n_sim}), or sometimes uniform rings but incorrect relative intensities of peaks as a function of $Q$.
As a result, the intensity of the peaks in the azimuthally integrated, $I(Q)$ are different to the $I(Q)$ of a perfect powder sample, though the peak positions are not altered. If this 1D pattern is handled as if it were from a perfect powder it results in a signal that is distorted from the \ppdf (\fig{multipanel_pdf}). This makes, depending on the extent of the texture, modeling of such \ppdf impossible (\fig{multipanel_pdf}).


The 3D PDF \cite{egami_underneath_2012} is a 3D histogram of atomic pair correlations separated by vector $\vec{r}$, rather than by distance $r$. Previously, the 3D PDF has been used to study diffuse scattering in single crystals \cite{weber_three-dimensional_2012, krogstad_reciprocal_2020}.  Similar to the case of the \ppdf, the 3D PDF of textured polycrystalline materials, \tpdf may be obtained by propagating the 3D total scattering structure function $S(\vec{Q})$ through the Fourier transform which is now a 3D transform,
\begin{equation}
	\label{eq:3DPDF}
	G(\vec{r}) = \frac{1}{(2\pi) ^{3}}
	\int [ S(\vec{Q}) - 1 ] e^{i\vec{Qr}} \: \dd \vec{Q},
\end{equation}
where
\begin{equation}
	\label{eq:iofqtosofq}
	S(\vec{Q}) = \dfrac{I(\vec{Q})}{\langle f(\vec{Q}) \rangle^{2}}
\end{equation}
is the 3D scattering function.   $S(\vec{Q})$ can be determined from a 3D scattering volume $I(\vec{Q})$, which can be reconstructed from a set of diffraction images.

Since a diffraction image measured on a 2D area detector is a slice on the surface of the Ewald sphere through reciprocal space, it is possible to reconstruct a 3D reciprocal scattering volume, $I(\vec{Q})$, from a sufficiently complete set of diffraction images with the sample at different orientations and from there obtain the 3D PDF according to Eq.~\ref{eq:3DPDF}.


The \tpdf formalism describes the texture by introducing the orientation distribution function (ODF) to a polycrystalline 3D PDF \cite{gong_atomic_2018}.  The ODF gives the probability density for a crystallite to face orientation $\vec{\Omega}$, which is the three-angle containing the Euler angles of the crystallite in the sample reference frame.



To yield the 3D PDF of a textured polycrystalline sample, \tpdf $G_{p}(\vec{r})$, a textured polycrystalline 3D structure function $S_{p}(\vec{Q})$ is first defined \cite{gong_atomic_2018} by integrating over all single crystallite structure functions $S'(\vec{Q},\vec{\Omega})$ at different orientation angles $\vec{\Omega}$, weighted by the ODF $D(\vec{\Omega})$,
\begin{equation}
	\label{eq:SpofQ}
	S_{p}(\vec{Q}) =
	\int D(\vec{\Omega}) S'(\vec{Q},\vec{\Omega}) \: \dd \vec{\Omega},
\end{equation}
where
\begin{equation}%
	\label{eq:S'ofQ}
	S'(\vec{Q},\vec{\Omega}) = 1 + \frac{1}{N' \langle f \rangle^{2}}
	\sum_{\substack{i \neq j}} f_{i}^{*}(Q) f_{j}(Q)
	e^{i\vec{Q} (\vec{R}({\vec{\Omega}})\vec{r}_{ij})}.
\end{equation}
Here $N'$ is the number of atoms in the crystallite, $\vec{r}_{ij}$ is the vector from atom $i$ to atom $j$, $\vec{R}({\vec{\Omega}})$ is the rotation matrix rotating $\vec{r}_{ij}$ from the sample reference frame to the respective crystallite orientation and $f_{i}$, $f_{j}$ are the atomic form factors of atom $i$ and $j$, respectively, in the crystallite. These equations are derived more fully in \cite{gong_atomic_2018}.
	
The \tpdf is then obtained by propagating $S_{p}(\vec{Q})$ through the Fourier transform,
\begin{equation}
	\label{eq:Gpofr}
	G_{p}(\vec{r})
	=  \frac{1}{(2\pi) ^{3}} \int [ S_{p}(\vec{Q}) - 1 ] e^{i\vec{Qr}} \: \dd \vec{Q}.
\end{equation}
	

\section{Example: A nanostructured thin film with fiber texture}
	
	

\begin{figure}
		\includegraphics[width=0.6\columnwidth]{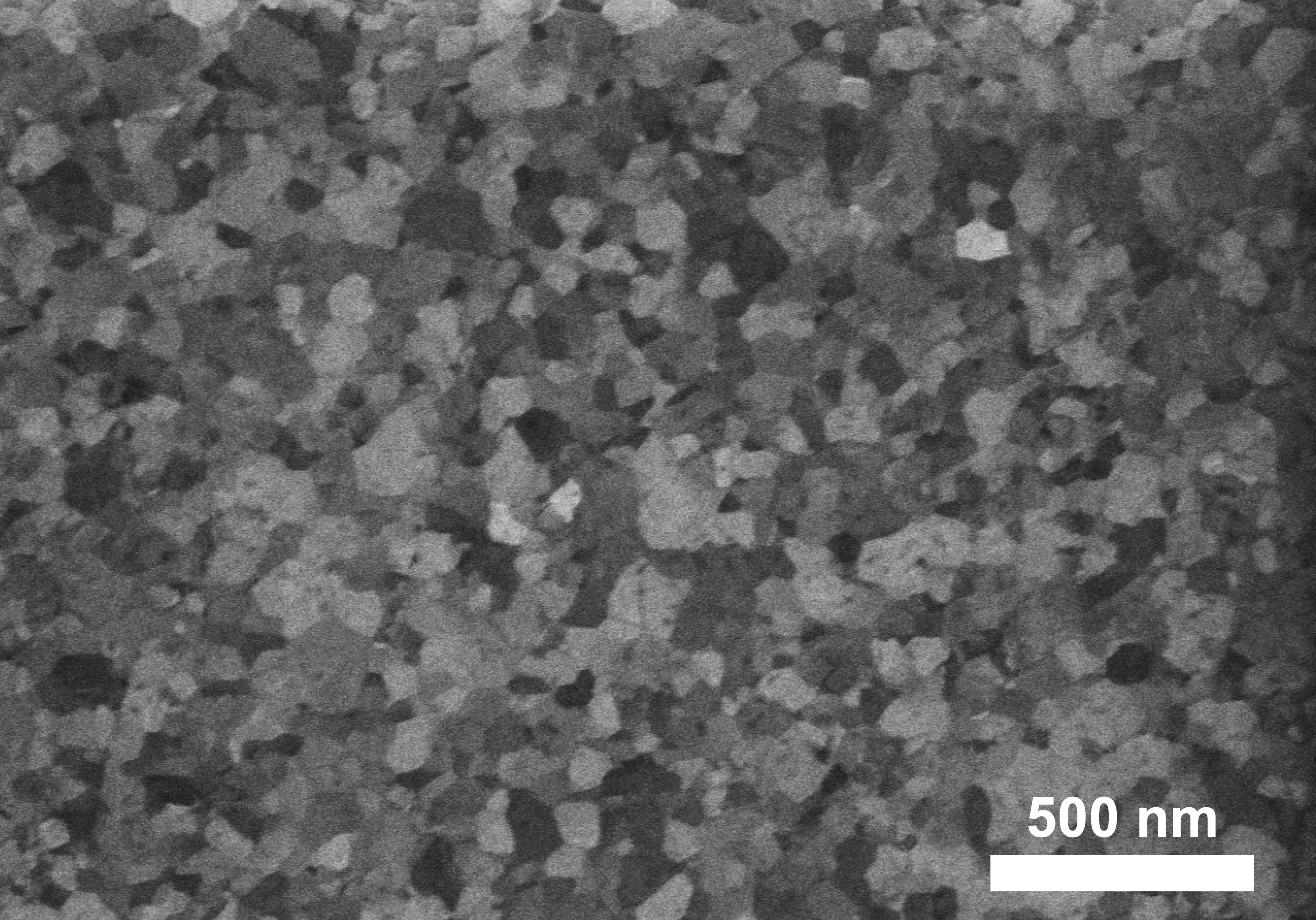}
		\caption{ SEM image of the Pt thin film surface showing a polydisperse size distribution and different shapes of the Pt grains in lateral direction.}
		\label{fig:sem}
	\end{figure}

Here we test the \tpdf method on experimental data for the case of a nanostructured polycrystalline Pt thin film that exhibits a strong fiber texture with $\langle 111\rangle$ crystallographic directions tending to align perpendicular to the film direction.  The sample was fabricated by sputter deposition (CS 500~ES, Von Ardenne GmbH) of a 50~nm platinum layer (target from FHR Anlagenbau GmbH) on a glass substrate with an intermediate evaporated 15~nm aluminum oxide adhesion layer (granulate from Evochem GmbH). An SEM image of the film is shown in Fig.~\ref{fig:sem}.
%


%
	\begin{figure}
		\includegraphics[width=0.99\textwidth]{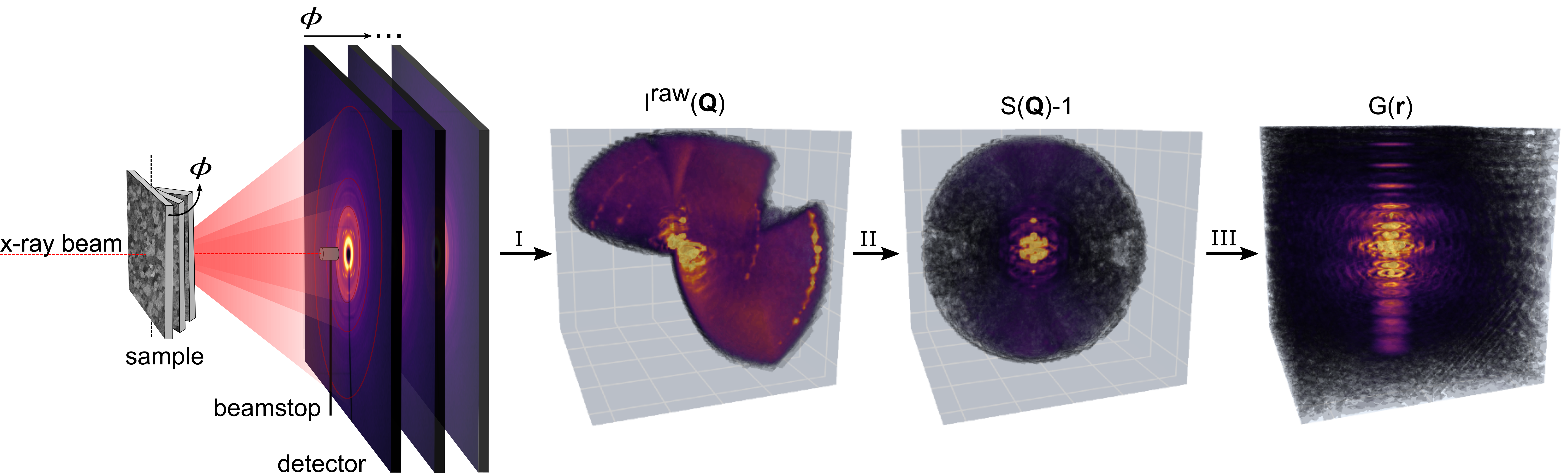}
		\caption{ Experimental and data processing procedure. $\textrm{I}$: Reconstruction of a tilt series of 2D scattering images to a 3D scattering volume $I(\vec{Q})$; $\textrm{II}$: Processing to the total scattering structure function $\textrm{S}(\vec{Q})$, application of symmetry averaging, interpolation and cut of beyond $\vec{Q}_\textrm{min}$ and $\vec{Q}_\textrm{max}$; $\textrm{III}$: Fourier transformation to the \tpdf $\textrm{G}(\vec{r})$.}
		\label{fig:analysis_procedure}
	\end{figure}
\fig{analysis_procedure} summarizes the experimental procedure and data processing.
The measurement was performed at beamline P07 at PERTRA III, Deutsches Elektronen-Synchrotron DESY, Hamburg, Germany \cite{gustafson_high-energy_2014}. The beam energy was 79.5~keV . The beam size was $0.5\times 0.5$~mm$^{2}$. Data was collected on a PerkinElmer XRD1621 detector.  The detector was calibrated by diffraction images of a lanthanum hexaboride coated glass substrate using the pyFAI software \cite{ashiotis_pyfai_2015}.

Diffraction images were measured at angles from $\phi = 1$$^\circ$ to 89$^\circ$ with a step size of $1^\circ$, where $\phi$ is the angle between the incident beam and the plane of the film.
First, the set of diffraction images are transformed from detector coordinates to reciprocal space coordinates using the python library meerkat \cite{simonov_meerkat_2019}.  Meerkat loads the diffraction images and bins each pixel to points in reciprocal space to reconstruct the 3D scattering volume.

	

Next it is necessary to make corrections to the data to convert raw intensities to the properly corrected and normalized $S(\vec{Q})$ function that can be Fourier transformed to the \tpdf.  
The measured scattering intensity from a sample will change during the measurement due to various experimental reasons, for example, the illuminated volume of the sample at different tilt angles, changes in incident beam intensity, and so on,  and this needs to be normalized.  If the geometry of the sample is well known, the incident beam monitored, and so on, these can be explicitly corrected~\cite{egami_underneath_2012}.  In this experiment the main intensity variation originates from the varying tilt angle.  In general, to account for such variations, we can apply a {\it post hoc} correction for these effects by applying a systematic scaling to the diffraction pattern as a whole. To determine the {\it post hoc} scaling factor we did the following.  First, we assume that away from the $Q$-values where Pt Bragg reflections exist, the scattering intensity is coming predominantly from the glass substrate and other incoherent sources (such as Compton scattering and any fluorescence) all of which are isotropic and should scale with the illuminated sample volume.
In this case, the scaling can be obtained from the total sum of the intensity in the azimuthally averaged $I^\prime(Q)$, i.e., $\sum_{i=1}^{N}I(Q_i)$, where the prime denotes that this $I^\prime(Q)$ excludes regions that contain Pt Bragg reflection intensities and the sum is over N intensity bins.  The value of this scale factor vs. $\phi$ is shown in Fig.~\ref{fig:scale_comparison}.  It is compared to the $1/\sin(\phi)$ scaling that is expected for our sample geometry~\cite{egami_underneath_2012}.  The agreement is very good except in the region of glancing angle, where the $1/\sin(\phi)$ diverges but the measured intensity from the sample turns over and comes back down.  This could be because of scattering from the edges of the sample, imperfect alignment of the sample and the detector, or due to the beam footprint exceeding the size of the sample.  If the sample is even a tiny bit too high or too low at grazing incidence the beam can partially miss the sample or be predominantly absorbed by the substrate, respectively, which would cause the observed effect.
These factors make grazing incidence (GI) experiments rather specialized in general~\cite{feidenhansl_surface_1989, dosch_critical_1992, gustafson_high-energy_2014}.  Here we are not concerned about making a high quality GI PDF measurement, but rather measuring the film over  a wide angular range, and the {\it post hoc} correction allows us to make use of the scattering deeper into the grazing incidence regime even with a non-ideal grazing incidence alignment and set up there.

\begin{figure}
	\includegraphics[width=0.9\columnwidth]{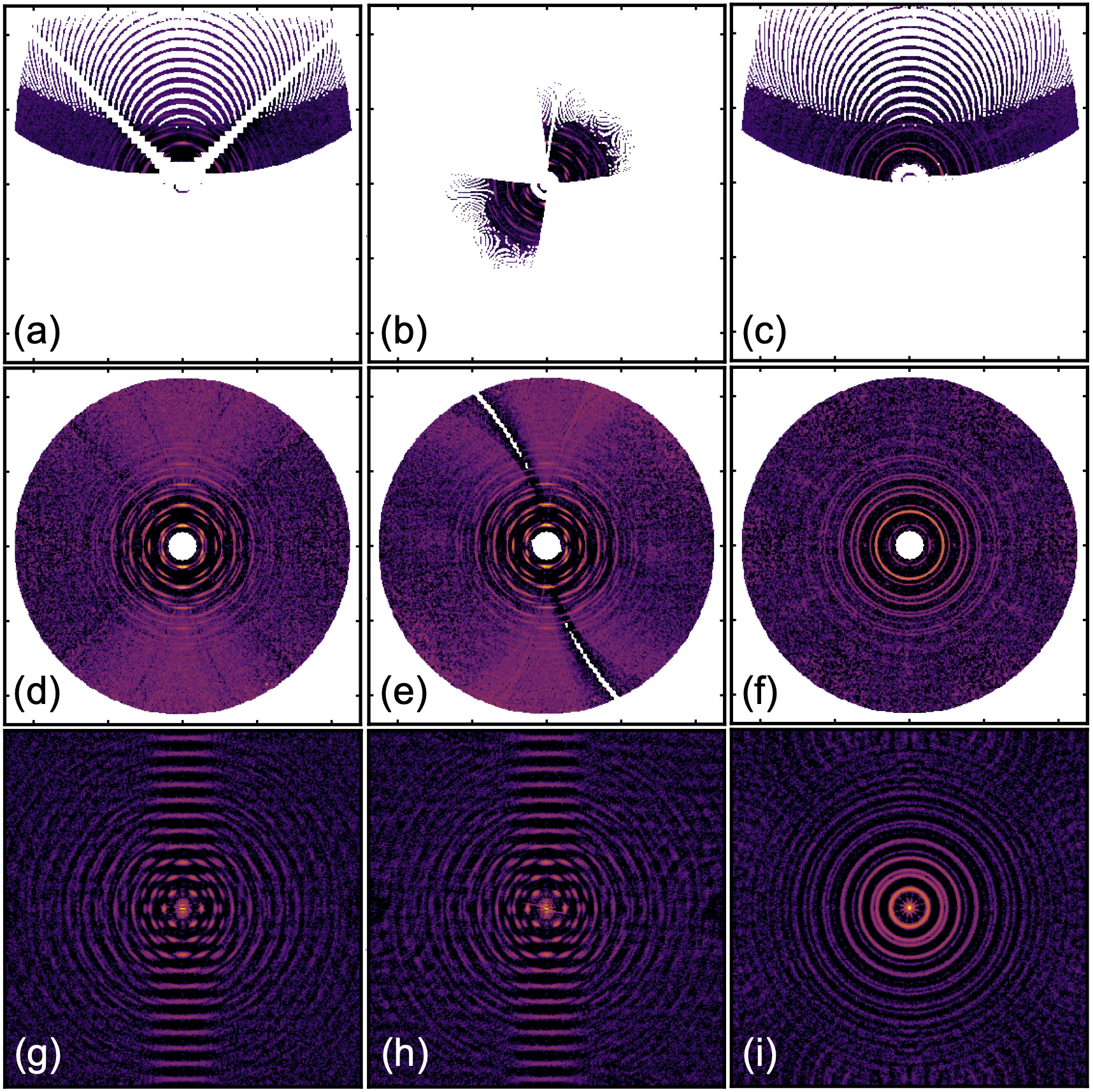}
	\caption{ Slices through the raw reconstructed scattering volume (a-c), processed total scattering structure function (d-f) and the \tpdf (g-i) with arbitrary units on a logarithmic scale. The slices in the first column are perpendicular to $x_s$, in the second column perpendicular to $y_s$ and in the third column perpendicular to $z_s$. All slices are through the origin of diffraction space. The edge length is from -23.9 to 23.9~\AA$^{-1}$ in (a-f) and from -26.3 to 26.3~\AA\, in (g-i), respectively. The void features in (a-c) result from the relatively high tilt step size of the measurement. The straight void lines in (a) result from masked high absorption features from the substrate. The void lines in (b) and (e) are due to the masking of the beamstop.}
	\label{fig:slices}
\end{figure}

	
Representative slices through reciprocal space of the  normalized, transformed, intensity data are shown in \fig{slices}(a-c).

There are large regions of reciprocal space missing.  This is because of the rather small tilt range and relatively high tilt step size in our experiment.  This can be remedied in general by taking many more images on a finer tilt grid, measuring over a wider angular range, and ultimately doing this around two perpendicular rotation axes.  However, in the current case, the high symmetry of the structure and the texture allow us to reconstruct the reciprocal space rather completely even from this limited set of data, through symmetrization.

In the figure, and below, we define the film plane normal to be the sample-$z$ direction, $z_s$, and select two perpendicular directions in the plane of the film to be the sample-$x$, $x_s$, and the sample-$y$ direction, $y_s$. We expect that referenced to these sample directions, the fiber texture dictates that the scattering will look the same viewed along the $x_s$ and $y_s$ directions, and look different, but circularly symmetric when viewed along $z_s$. If we take the actual diffraction images and reconstruct them into this 3D diffraction space,  and make slices through the reconstructed diffraction space perpendicular to $x_s$, $y_s$, and $z_s$, respectively, going through the center of reciprocal space we get the images in Fig.~\ref{fig:slices}(a-f).  It is difficult to verify the predictions in the unsymmetrized data (Fig.~\ref{fig:slices}(a-c)), but is very apparent after filling in the missing regions (Fig.~\ref{fig:slices}(d-f)).  For this fiber texture, the symmetrization consists simply of rotating about the fiber axis and averaging the intensities.

The symmetrized diffraction signal is then converted to $S(\vec Q)-1$ by dividing by the average form factor squared and rescaling, as shown in \fig{analysis_procedure}.  In practice, the measured intensity comprises several features in addition to the coherent scattering intensity such as incoherent Compton scattering, multiple scattering, self absorption and polarization effects of the x-ray beam, which we should correct for before Fourier transforming to the PDF \cite{egami_underneath_2012}.  In case of the \ppdf the scattering intensity can be corrected by modeling a polynomial correction term $\beta(Q)$ \cite{juhas_pdfgetx3_2013}.  By assuming the scattering contributions are isotropic, we can calculate $\beta(Q)$ by propagating the spherical integration of $S(\vec Q)-1$, $S(Q)-1$ which should resemble the powder average, through the \textit{PDFgetX3} algorithm \cite{juhas_pdfgetx3_2013}.  We then isotropically subtract the polynomial correction $\beta(\vec Q)$ from $S(\vec Q)-1$ to retrieve the coherent scattering fraction.  The same slices as shown in \fig{slices}(a-c) symmetrized and in the form of $S(\vec Q)-1$ are shown in \fig{slices}(d-f). The 3D diffraction intensities from a perfect powder would appear as concentric spheres in this reconstructed scattering volume, and so slices through the volume that go through the origin of reciprocal space, as we are plotting in the figure, would show uniform circles.  This is indeed what we see in the slice looking down $z_s$, the fiber axis (\fig{slices}(f)).  The circles are also evident in the slices perpendicular $x_s$ and $y_s$ ((d) and (e)).  However, the circles appear broken and have non-uniform intensities as you go around the azimuth as a result of the texture.  The two slices are, of course, identical to each other due to the fiber symmetry and the resulting symmetrization step.
	

The scattering volume is further interpolated and intensity values beyond $Q_{\textrm{min}}$ and $Q_{\textrm{max}}$ are set to zero before carrying out the 3D Fourier transformation to obtain the \tpdf\ $G(\vec{r})$ (Eq.~\ref{eq:3DPDF}), shown in \fig{analysis_procedure}, and the slices shown in \fig{slices}(g-i).  The spherical symmetry of the scattering from a perfect powder also results in a spherically symmetric $G(\vec{r})$.  In the false-color image in  \fig{slices}(g-i) we would expect to again see circular patterns of increased and reduced $G$ values.  These circles are evident when viewed down the $z_s$ axis (\fig{slices}(i)), but as before become non-uniform around the azimuth when viewed along the $x_s$ and $y_s$ axes (\fig{slices}(g,h)).


\section{Texture Analysis}
	
	
	
	
	
	The reconstructed reciprocal space, and resulting real space \tpdf, presented above contain concentric spheres of intensity at values of reciprocal lattice vector magnitude, $|\vec{H}^{hkl}|$ (reciprocal space) and interatomic vector magnitude, $|\vec{r}^{i,j}|$ (real-space \tpdf), that become broken up and  non-uniform due to the presence of texture.  There is, overall, a resemblance in the patterns of intensity between $G(\vec{r})$ and those seen in reciprocal space as they contain the same fiber texture information, but peaks in $G(\vec{r})$ appear at positions corresponding to interatomic vectors rather than reciprocal lattice vectors.  We now consider how to extract texture information from $G(\vec{r})$.
	
	
	
\begin{figure}
		\centering
		\includegraphics[width=0.9\columnwidth]{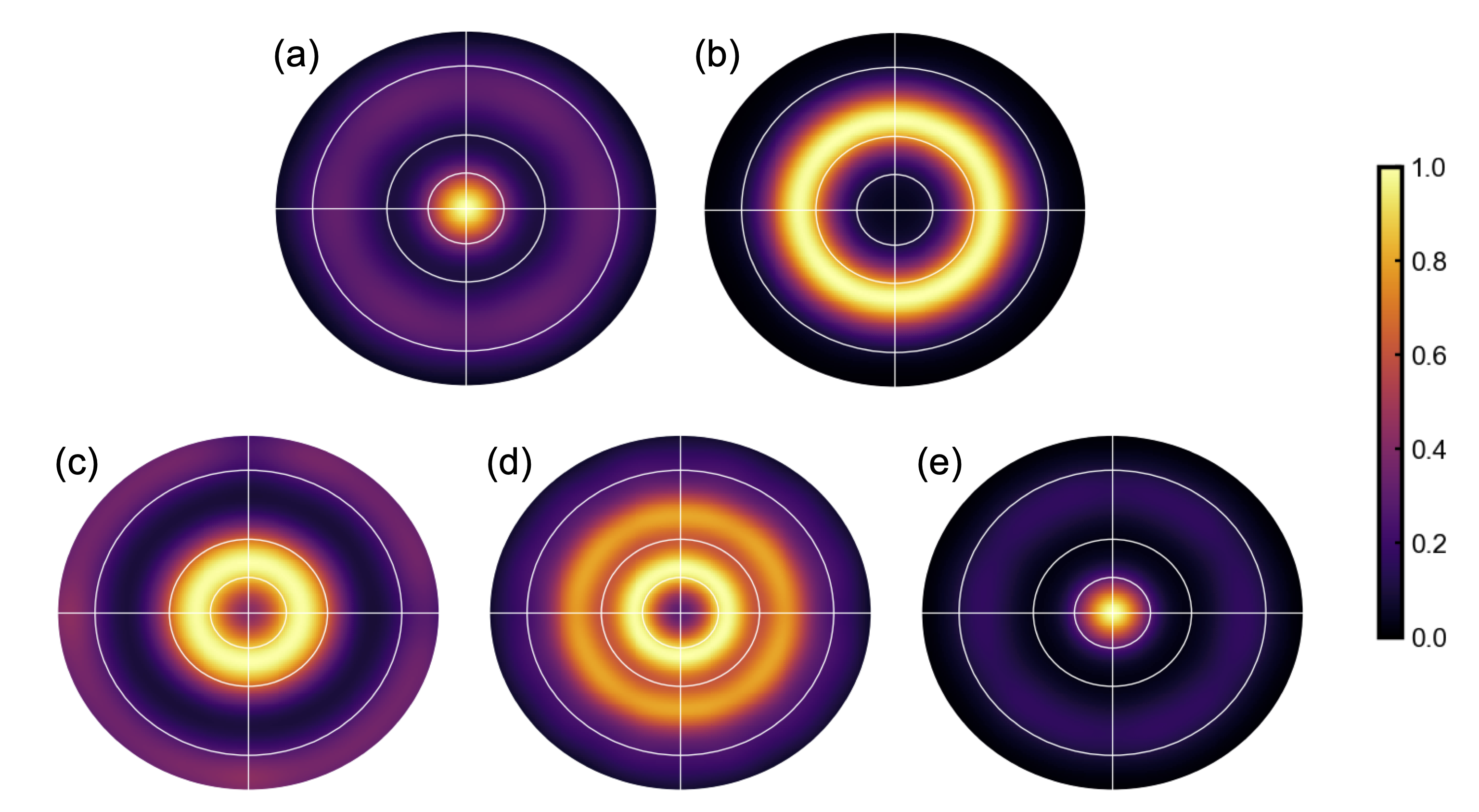}
		\caption{Reciprocal space pole figures derived from the scattering volume $S(\vec{Q})$. Peaks: (a): \{111\}, (b): \{002\}, (c): \{022\}, (d): \{133\}, (e): \{222\}.}
		\label{fig:polefigures_reci}
	\end{figure}

	One way to understand the crystalline texture is to use the pole figure construction~\cite{bunge_texture_1982}.  In this construction we consider a reciprocal lattice vector from a single crystal grain, $\vec{H}_1^{\textrm{hkl}}$, and place its tail at the origin of our diffraction space.  We then consider this same reciprocal lattice vector from every crystal grain, $m$, in the sample and do the same.  This results in a construction where the tip of all the reciprocal lattice vectors, $\vec{H}_m^{\textrm{hkl}}$, lie on the surface of a sphere of radius $\vert \vec{H}^{\textrm{hkl}} \vert$, which we here refer to as the reciprocal space pole sphere.  In a perfect powder there would be a uniform density of vector tips over the entire surface of the reciprocal pole sphere.  When texture is present, some regions of the surface of the reciprocal space pole sphere are covered more densely than others.  In the case of a fiber texture such as ours, it is conventional to align the north pole of the reciprocal space pole sphere with the sample fiber axis.  In this case we find high densities of pole vectors that form spherical segments (or spherical caps for the $\vec{H}_m^{\textrm{111}}$) on the pole sphere.  The pole figure is created by taking a stereographic projection of the vector tips on the reciprocal space pole sphere.  In practice, for a given set of reciprocal lattice vectors, $\vec{H}_m^{\textrm{hkl}}$, the intensity of scattering on a spherical annulus centered at $Q=\vert \vec{H}^{\textrm{hkl}} \vert$ in the reconstructed diffraction space is proportional to this quantity (assuming there are no other overlapping reciprocal lattice vectors).  The stereographic projection of these intensities results in the pole figure for that particular reciprocal lattice vector, or set of symmetry equivalent reciprocal lattice vectors.  In the pole figure the fiber texture appears as circular rings of bright intensity as shown in \fig{polefigures_reci}.
	
\begin{figure}
		\centering
		\includegraphics[width=0.9\columnwidth]{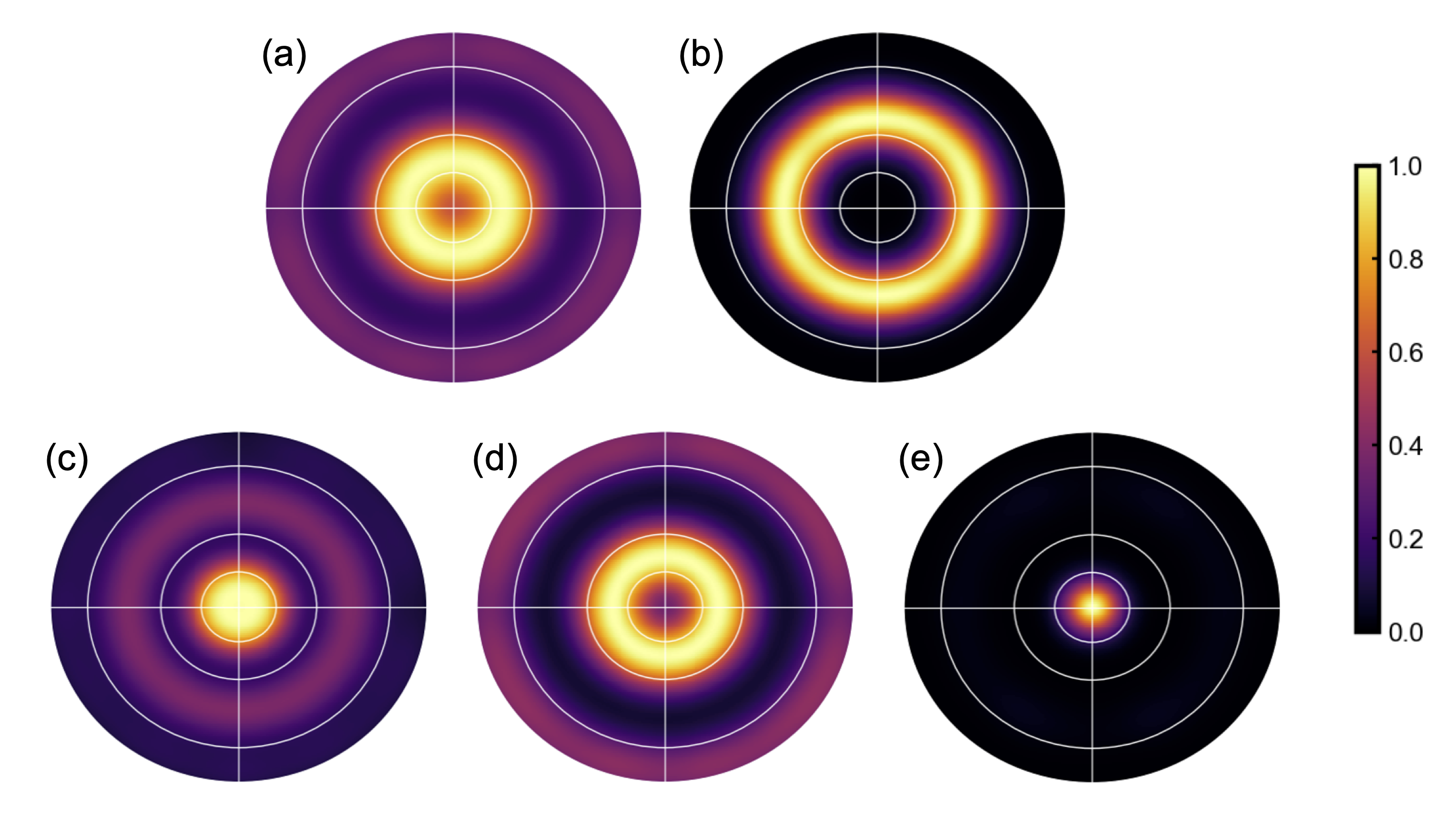}
		\caption{Real space pole figures derived from the \tpdf $G(\vec{r})$. Peaks: (a): $ 0, \frac{1}{2}, \frac{1}{2} $, (b): $ 0,0,1 $, (c): $ \frac{1}{2}, \frac{1}{2}, 1 $, (d): $ 0,1,1 $, (e): $ 1,1,1 $.}
		\label{fig:polefigures_real}
	\end{figure}

	
	The standard pole figure construction is inaccurate for materials that are not long range-ordered since the scattering is not concentrated in sharp Bragg peaks lying at the reciprocal lattice points.  However, as shown above, it is possible to describe texture in terms of real-space structural vectors through the 3D \tpdf construction.  We therefore explore the real space analog of the pole figure as a stereographic projection of the intensity on the surface of the real space pole sphere which is defined analogously to the reciprocal space pole sphere but for interatomic vectors $\vec{r}^{i,j}$.  We call the resulting stereograms ``real space pole figures".
	Projecting on the equatorial plane perpendicular to the fiber axis $z_s$ yields, as in the reciprocal space pole figure, uniform concentric rings of bright intensity. Representative real space pole figures from our sample are shown in \fig{polefigures_real}.
	
	
	
\begin{figure}
		\includegraphics[width=0.99\columnwidth]{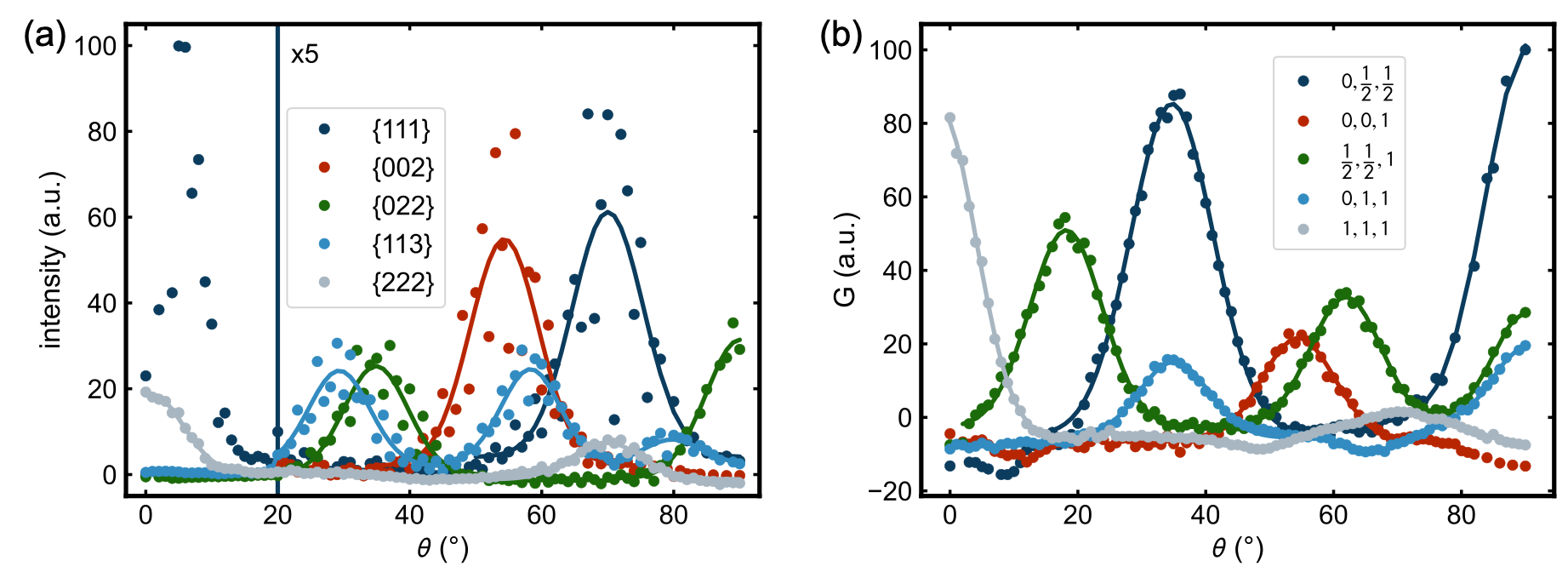}
		\caption{ (a) Reciprocal space fiber plots of the first five XRD reflections.  (b) Real space fiber plots of the first five PDF peaks.  Gaussian fits of the peaks in the fiber plots are shown as solid traces. The fit of the reciprocal 111 reflection at $\theta$ = 0$^\circ$ is omitted because of low data quality at that region and poor resolution at low Q after the conversion to polar coordinates.  The fit of the real 111 peak at $\theta$ = 70.53$^\circ$ is omitted because of a poor fit result due to the low intensity.  The intensities of the reciprocal and real space fiber plots are retrieved by summing over all intensity values at the $H^{hkl}$ value of the respective peak ± 0.1 $\mathrm{\AA}^{-1}$ from the scattering volume $S(\vec{Q})$ and over the $r^{ij}$ value  of the respective interatomic distance ± 0.1~\AA\, from the \tpdf $G(\vec{r})$, respectively. Increasing the integration width for the fiber plot construction does not reduce the high noise in the reciprocal space fiber plot.}
		\label{fig:fiberplots}
	\end{figure}

	Due to the fiber texture and the choice of projection plane, the pole figure intensities are independent of the azimuthal angle and we can integrate them azimuthally to obtain 1D plots called fiber plots \cite{bunge_texture_1982, he_two-dimensional_2018}.  In the fiber plot representation the intensity is plotted against the zenith (i.e. polar) angle, $\theta$, which is the angle between the position vector of the pixel on the real/diffraction-space spheres and the polar axis, $z_s$. In the center of the pole figure $\theta$ is $0^{\circ}$ and the outer ring at the edge of the pole figure is at $\theta = 90^{\circ}$.  The reciprocal and real space fiber plots corresponding to these pole figures are shown in \fig{fiberplots}, respectively.
	
	
	Assuming the fiber texture gives a Gaussian distribution of poles as a function of the zenith angle $\theta$,
	we can fit a series of Gaussian functions to the peaks in the fiber plots. Apart from some noise in the data we get stable fits.  The positions of the fitted peaks in the reciprocal as well as the real space fiber plots, $\theta^{H^{hkl}}_{0}$ and $\theta^{r^{ij}}_{0}$, respectively, are in good agreement with the expected angle between the respective reciprocal/real vector and the 111 axis, which is the fiber axis  (see Table \ref{tab:reci_fiberplot_fits} and \ref{tab:real_fiberplot_fits}).  The width of the peaks in the fiber plot comes from the degree/extent of the fiber texture.  We get similar results for the full width at half maximum (FWHM) in the reciprocal and the real fiber plots, which confirms that our treatment of the data to obtain the \tpdf is valid.  We do notice that the real FWHMs, at around $13.6^\circ \pm 2.1^\circ$, are slightly larger than the reciprocal FWHMs ($11.6^\circ \pm 1.2^\circ$), where the uncertainties represent the spread in values obtained from all the different reciprocal/real vectors.  Although the difference between the real and reciprocal measures is within the variation of values the \tpdf is giving a width for the fiber texture that is rather consistently larger than that obtained from the reciprocal data.  We think that this difference is real and may reflect the presence of mosaicity in the crystals that would be seen as a distribution in interatomic vector angles within a single crystallite and seen in the \tpdf but not in the reciprocal space pole figures and fiber plots.  We also note from Fig.~\ref{fig:fiberplots} that the scatter of the data points is much greater in the reciprocal-space fiber plots than the real space ones. We are uncertain as to the origin of this effect, but for the real-space plots it means that the variation of FWHM values is larger than any variation due to the noise in the data.  This may also indicate that it is possible to measure a directionality to the mosaicity within the crystals resulting in small variations in width in different  crystallographic directions, though a full analysis of this is beyond the scope of the current paper.

	
	Although, in the case of the fiber texture, pole figures and fiber plots yield a straightforward and rather precise picture of the texture, in general the texture of a sample cannot be comprehensively described by the two angles of a pole figure.  The more complete texture representation is the ODF, $D(\vec{\Omega})$, which describes the crystallite orientation by the three Euler angles in a density probability function \cite{bunge_texture_1982}.  The ODF can be evaluated from a series of pole figures, where the number needed depends on the symmetry of the texture and of the crystal \cite{bunge_texture_1982}.
	However, in general the ODF may be reconstructed directly by modeling the 3D structure function $S(\vec{Q})$ or \tpdf $G(\vec{r})$ \cite{gong_atomic_2018}.  If the structure function of the reference crystallite $S'(\vec{Q})$ (eq.~\ref{eq:S'ofQ}) is known, as is often the case, we can rotate $S'(\vec{Q})$ at different orientations of $\vec{Q}$ and add the contributions weighted by an initial assumption of the ODF to build the polycrystalline structure function $S_p(\vec{Q})$ (eq.~\ref{eq:SpofQ}). Now we can fit $S_p(\vec{Q})$ to the measured $S(\vec{Q})$ by modeling the ODF in a regression loop.  Equivalently, the ODF can be retrieved by a similar modeling of the \tpdf $G(\vec{r})$ (eq.~\ref{eq:Gpofr}) \cite{gong_atomic_2018}.
	
	
	\begin{table}
		\caption{Reciprocal space fiber plot peak fit results. $\theta^{\vec{H}^{hkl}}_{0}$ is the fitted peak position.}
		\label{tab:reci_fiberplot_fits}
		\centering
		\begin{tabular}{cccc}
			\hline
			lattice plane & \multirow{2}{2.5cm}{\centering reference angle \linebreak to [111] ($^{\circ}$)} & \multicolumn{2}{c}{Gaussian fit result}\\
			
			& & $\theta^{\vec{H}^{hkl}}_{0}$($^{\circ}$) & FWHM($^{\circ}$)\\
			\hline
			\{111\} & $\;\:$0.0 & - & - \\
			& 70.5 & 70.0(6) & 13(2)\\
			
			\{002\} & 54.7 & 54.3(5) & 12.2(14)\\
			\{022\} & 35.3 & 34.9(3) & 11.4(13)\\
			
			& 90.0 & 89.8(7) & 12(2)\\
			
			\{113\} & 29.5 & 29.3(3) & 10.9(12)\\
			
			& 58.5 & 58.3(3) & 10.4(14)\\
			
			& 80.0 & 79.8(6) & 10(3)\\
			
			\{222\} & $\;\:$0.0 & $\;\:$0.9(3) & 11.4(6)\\
			
			& 70.5 & 69.8(2) & 12.5(7)\\
			\hline
		\end{tabular}
	\end{table}
	\begin{table}
		\caption{Real space fiber plot peak fit results. The interatomic vectors from atom $i$ to atom $j$ are represented in terms of components using a basis of the crystallographic lattice vectors and assuming that the origin is placed on atom $i$.  $r^{ij}$ is the distance between atom $i$ and atom $j$.  $\theta^{r^{ij}}_{0}$ is the fitted peak position.}
		\label{tab:real_fiberplot_fits}
		\centering
		\begin{tabular}{ccccc}
			\hline
			\multirow{2}{1.5cm}{\centering position\linebreak of atom $j$} & $r^{ij}$ & \multirow{2}{2.5cm}{\centering reference angle\linebreak to [111]($^{\circ}$)} & \multicolumn{2}{c}{Gaussian fit result}\\
			
			& & & $\theta^{r^{ij}}_{0}$ ($^{\circ}$) & FWHM ($^{\circ}$)\\
			\hline
			$ 0, \frac{1}{2}, \frac{1}{2} $ & 2.77~\AA & 35.3 & 34.72(7) & 15.2(2)\\
			& & 90.0 & 90.5(6) & 15.7(10)\\
			
			$ 0,0,1 $ & 3.92~\AA & 54.7 & 54.47(10) & 13.5(3)\\
			
			$ \frac{1}{2}, \frac{1}{2}, 1 $ & 4.81~\AA& 19.5 & 18.16(9) & 13.4(3)\\
			& & 61.9 & 61.92(8) & 13.0(3)\\
			& & 90.0 & 90.2(6) & 11.8(13)\\
			
			$ 0,1,1 $ & 5.55~\AA& 35.3 & 35.04(10) & 12.8(3)\\
			& & 90.0 & 90.1(7) & 12(2)\\
			
			$ 1,1,1 $ & 6.80~\AA & $\;\:$0.0 & -1.5(4) & 13.3(6)\\
			& & 70.5 & - & - \\
			\hline
		\end{tabular}
	\end{table}
	
	
\section{Structural Analysis}
	
	
	
	Here we discuss extracting meaningful structural information from PDFs that were measured from textured samples.
	This is relevant for the case where the experimenter is not trying to characterize the sample texture per se, but is interested in the local structure of a textured sample.  In favorable circumstances, texture related intensity changes in Bragg peaks can be modelled in a Rietveld refinement \cite{gilmore_international_2019}.
	This is less well developed in real-space refinements, though for weak textures, a model has been proposed \cite{cervellino_texture_2020}.
	Here we explore whether sufficiently texture-free PDFs, which can then be refined in existing PDF modeling programs such as PDFgui \cite{farrow_pdffit2_2007}, can be obtained by averaging data collected at different sample angles.  As a test we will use the dataset from the fiber-textured Pt thin film as an example.
	
	
	In theory, measuring the diffraction pattern by sampling every orientation of the sample with equal statistics will result in a spherically homogeneous diffraction pattern that can be processed as an ideal powder pattern.  Using a 2D detector and rotating 180$^\circ$ about two mutually perpendicular axes perpendicular to the beam can accomplish this for any crystallographic and texture symmetry, if correct counting statistics are ensured.  Correct counting statistics will depend on flux and exposure time weighting, but also any geometrical and absorption effects that result in changes in illuminated sample volume.   These can be readily controlled for simple sample shapes (spherical, cylindrical, planar/film, etc.) as we describe below.  In many cases, such as the case of the fiber-textured sample, the averaging may be accomplished by rotating about a single axis perpendicular to the beam, provided the fiber axis doesn't lie along the rotation axis.  For experimental setups where this is possible, and where data-rates are sufficiently high, it probably represents the best approach.   In our case the fiber texture resulted from oriented growth of the polycrystalline film on the substrate.  In this case the illuminated volume is well described by the area of the beam footprint on the sample, which is given by the beam width, $w$, times $h/\cos\theta$, where $h$ is the beam height and $\theta$ is the angle the substrate plane normal makes with the incident beam. Since $w$ and $h$ don't vary with incident angle, we get that the measured intensities should be multiplied by   $1/\cos(\theta)=1/\sin(\phi)$ \cite{egami_underneath_2012} where $\phi = \pi/2 - \theta$ is the grazing angle the incident beam makes with the substrate.   For even higher accuracy, we use the same normalization/scaling approach which is applied to the diffraction images before reconstructing to the 3D scattering volume as discussed in the experimental section.
	
	\fig{multipanel_pdf}(e) shows a fit of an untextured model to the PDF that was obtained by a properly weighted average of the PDFs from all sample angles collected in our dataset.  Indeed the fit is superior to the fits at any fixed sample angle, as evident from the difference curve.  Table \ref{tab:1dpdf} shows the refined structural parameters and the goodness of the fit parameter, $R_{w}$ of a few single-angle PDF refinements and the averaged dataset.
	\begin{table}
		\caption{1D PDF refinement results to the Pt thin film measured at different incidence angles, $\phi$, and a fit to an experimental PDF obtained as an average over all tilt angles. Refinement parameters: "$a$" is the lattice parameter,  $U_{iso}$ is the isotropic atomic displacement parameter (ADP), $\delta_{2}$ is the coherent thermal motion parameter, and $R_{w}$ is the goodness of the fit.}
		\label{tab:1dpdf}
		\centering
		\begin{tabular}{cccccc}
			\hline
			$\phi$ ($^{\circ}$) & a ($\textrm{\AA}$) & ADP ($\textrm{\AA}^{2}$) & $\delta_{2}$ ($\textrm{\AA}^{2}$) & $R_{w}$ (\%)\\
			\hline
			1 & 3.924 & 0.00879 & 4.99 &  42.2 \\
			11 & 3.932 & 0.00538 & 3.66&  18.1 \\
			23 & 3.929 & 0.00503 & 4.25 &  20.0 \\
			45 & 3.929 & 0.00548 & 5.45 & 27.8 \\
			67 & 3.932 & 0.00518 & 6.24 &  35.7 \\
			89 & 3.933 & 0.00398 & 7.80 & 63.6 \\
			avg & 3.931 & 0.00550 & 3.93 & 12.0 \\
			\hline
		\end{tabular}
	\end{table}
	Indeed, the best fit of the untextured model to the data is obtained for averaged data ($R_w = 12$\%) compared to any single fixed angle dataset.  The best agreement between the powder average model and a single-angle dataset is achieved for $\phi = 11^\circ$ with $R_w = 18$\%. For this powder texture the agreement gets better as the incident angle gets shallower, though at small incident angles (below $\phi = 11^\circ$) the measured PDFs seem to become less reliable. We note that the refined parameters of the fit to the $\phi = 11^\circ$ are also closer to those obtained for the averaged data than for the other angles.  A plot of the best-fit $R_w$ vs. $\phi$ is shown in  Fig. \ref{fig:pdf_ref_res} showing this behavior in a more quantitative fashion.
	Based on this analysis, when looking at data from a textured thin-film sample, the best strategy for obtaining PDFs that represent a texture-less structure and are suitable for conventional quantitative structural modeling in real space it is recommended to rotate the sample $\pm 89^{\circ}$ about an axis perpendicular to the beam (from a starting position with the film perpendicular to the beam) and to average the data as described in this paper.  If this is not possible or proves inconvenient, for similarly textured films, taking a single dataset at an incident angle of $\sim 10^{\circ}$ grazing angle to the film will give a reasonably untextured signal.

	
	
	Beyond summing raw images from different angles, it is also possible to obtain an 1D powder pattern and PDF resembling that of an untextured sample by integrating spherical shells in the reconstructed \tpdf . In theory, independent of the kind, extent and orientation of the texture, this integration should resemble the precise isotropic powder average and can be processed as such. Correspondingly a similar integration of the reconstructed scattering volume can be analyzed as an isotropic powder pattern. We were not able to test this in the current case because our test data were not collected on a fine enough grid of angles to avoid interpolation errors that introduced significant aberrations.
	
	
	Finally, we note in passing that the \tpdf can also be applied to study other anisotropic structural properties such as strain and stress by analyzing the \tpdf along different directions in a small angular range.
	
	
	
\section{Software}
	
The software used to do the work in this paper has been turned into a python package and is being released under an open source license.  The package is called Fourigui and is available through the diffpy website (https://diffpy.org).
It is available on the python package index and on Anaconda through the conda-forge channel. The code is hosted on GitHub at https://github.com/diffpy/fourigui.

Fourigui loads a 3D total scattering volume, for example, obtained from a rotation image set, using a software package such as XCAVATE \cite{estermann_diffuse_1998}, XDS \cite{kabsch_xds_2010} or meerkat \cite{simonov_meerkat_2019}, and performs the FFT to the \tpdf with optional cutoff frequencies. It also incorporates visualization capabilities for viewing the data in real and reciprocal space.  It always displays one slice perpendicular to one of the three axis of either the scattering volume or the \tpdf obtained with or without cut off frequencies and allows easy scrolling through the full volume to visualize the intensity distribution.  \fig{screenshot_fourigui} provides further insight about the utilization of Fourigui.


	
\section{Conclusion}
	
	

Here we present a practical implementation of the 3D PDF method of textured polycrystalline sampels, \tpdf.  We demonstrate the procedure for measuring and analyzing data to get the \tpdf\ using a fiber textured Pt thin film sample.   2D diffraction images of the Pt thin film measured at different tilt angles are reconstructed to a scattering volume in 3D reciprocal space.  The scattering volume is then processed to the 3D total scattering structure function and propagated through the 3D Fourier transform to yield the \tpdf in real space. Further, we present a definition, and demonstration, of real space pole figures.  We also present Fourigui, a python package for carrying out the steps in the procedure.


The \tpdf method may be used to study texture in highly nanocrystalline samples where Bragg reflections are broad and overlapped preventing to study the texture directly in reciprocal space. The approach we present may also be used to obtain approximately isotropic 1D PDFs from textured samples, which will be useful when the structure of a sample is sought, rather than a measure of its texture, but the experimental data are textured.  A good isotropic PDF approximation was obtained from the weighted average of the diffraction images, yielding a reasonable PDF refinement, $R_{w} = 12\%$, from the averaged PDF compared to the single tilt angle PDFs with $R_{w}$ values ranging from 18\% to 64\%.  We also showed that an approximately ideal 1D PDF could be obtained from a not-quite grazing incidence (10$^\circ$ incident angle) single measurement in the case of a thin film with a fiber texture.



The \tpdf is very promising to study nanocrystalline thin films as they are likely to be textured and enable a straightforward measurement of the tilt series.  However, the \tpdf can be applied in general to materials of any geometry.  Further, we note that the methodology of the \tpdf can be expanded to study nanoscale materials with different structural anisotropies such as anisotropic strains.

	
	
	
\section*{Acknowledgements}
The authors would like to thank Zach Thatcher, Connor Bracy, Silvio Achilles and Ludwig Hendl for help with software releases.  Work on \tpdf developments and analysis in the Billinge group were supported by the U.S. Department of Energy, Office of Science, Office of Basic Energy Sciences (DOE-BES) under contract No. DE-SC0012704. SYHM and DK acknowledge support from the Cluster of Excellence 'CUI: Advanced Imaging of Matter' of the Deutsche Forschungsgemeinschaft (DFG) - EXC 2056 - project ID 390715994 and from the Bundesministerium für Bildung und Forschung (BMBF) via project LUCENT under grant No. 05K19WMA.
We acknowledge DESY (Hamburg, Germany), a member of the Helmholtz Association HGF, for the provision of experimental facilities. Parts of this research were carried out at PETRA~III.  We would like to thank Uta Ruett, Olof Gutowski und René Kirchhof for assistance in using P07. 
Further, we thank Daliborka Erdoglija, Fenja Berg, Dagmar Leisten and Ulrich Boettger of the Institute for Materials in Electrical Engineering (IWE-2), RWTH Aachen University (Aachen, Germany) for the fabrication of the textured Pt thin film.

\section{Supporting Information}
	\setcounter{figure}{0}
	\setcounter{table}{0}
	\makeatletter
	\renewcommand{\thetable}{S.\arabic{table}}
	\renewcommand{\thefigure}{S.\arabic{figure}}
	\makeatother
	
	\begin{figure}
		\includegraphics[width=0.9\columnwidth]{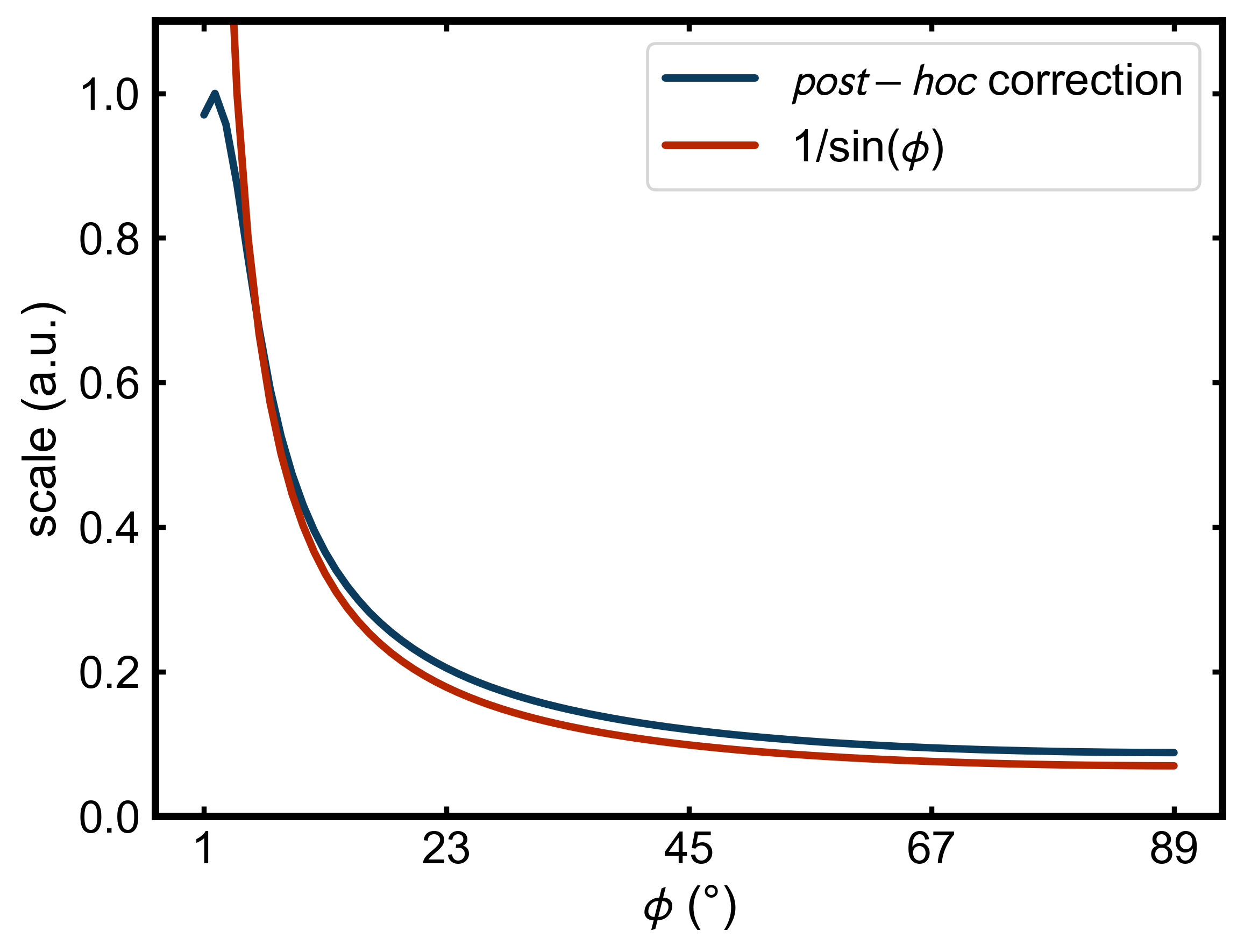}
		\caption{\textit{Post-hoc} correction versus mathematical correction for the intensity scaling of the diffraction images and patterns taking into account the different illuminated volumes of the sample at different tilt angles $\phi$.}
		\label{fig:scale_comparison}
	\end{figure}
	
	\begin{figure}
		\includegraphics[width=0.9\columnwidth]{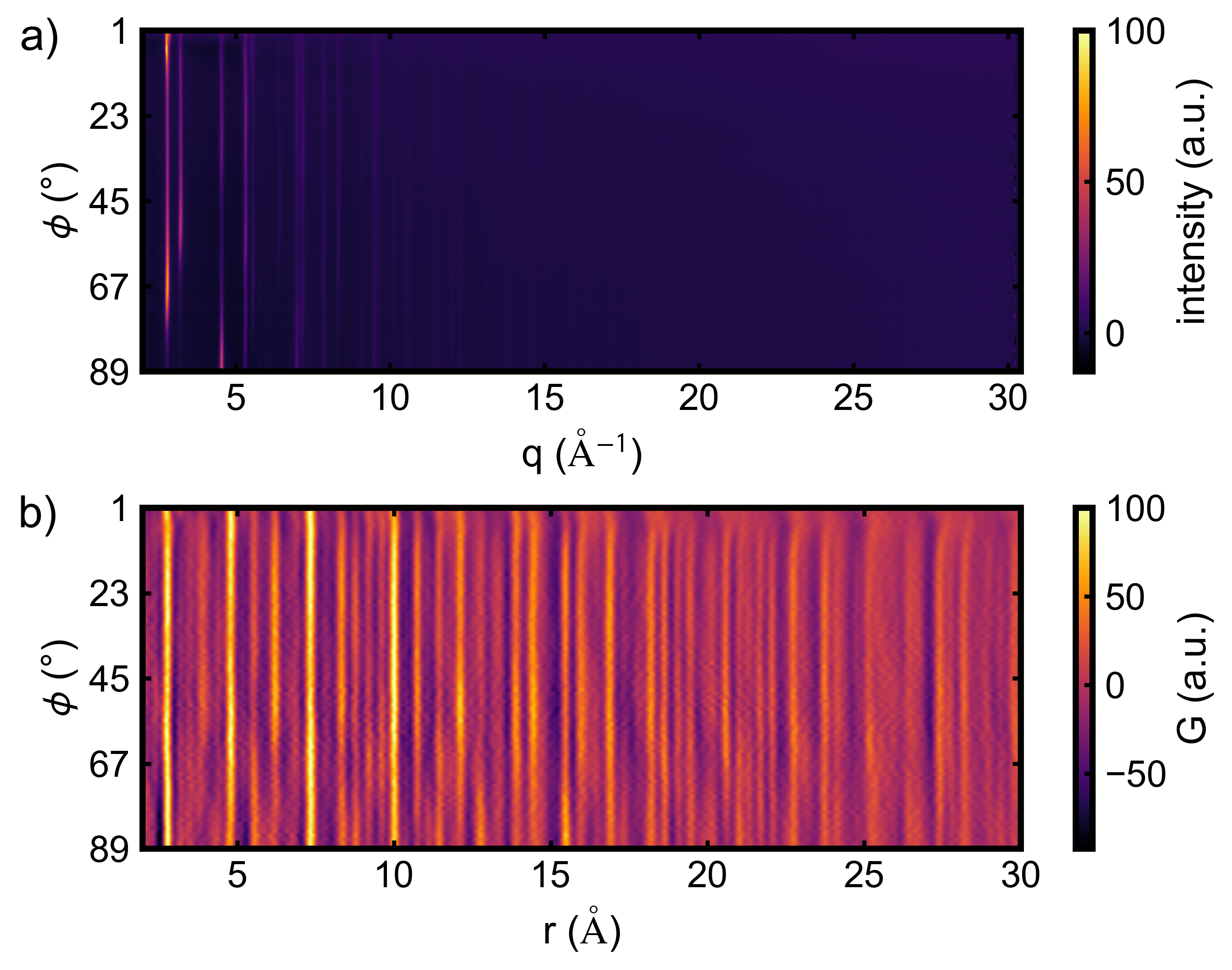}
		\caption{\label{fig:heatmap} a) 1D XRD and b) 1D PDF from the Pt thin film at different tilt angles $\phi$.}
	\end{figure}
	
	\begin{figure}
		\includegraphics[width=0.9\columnwidth]{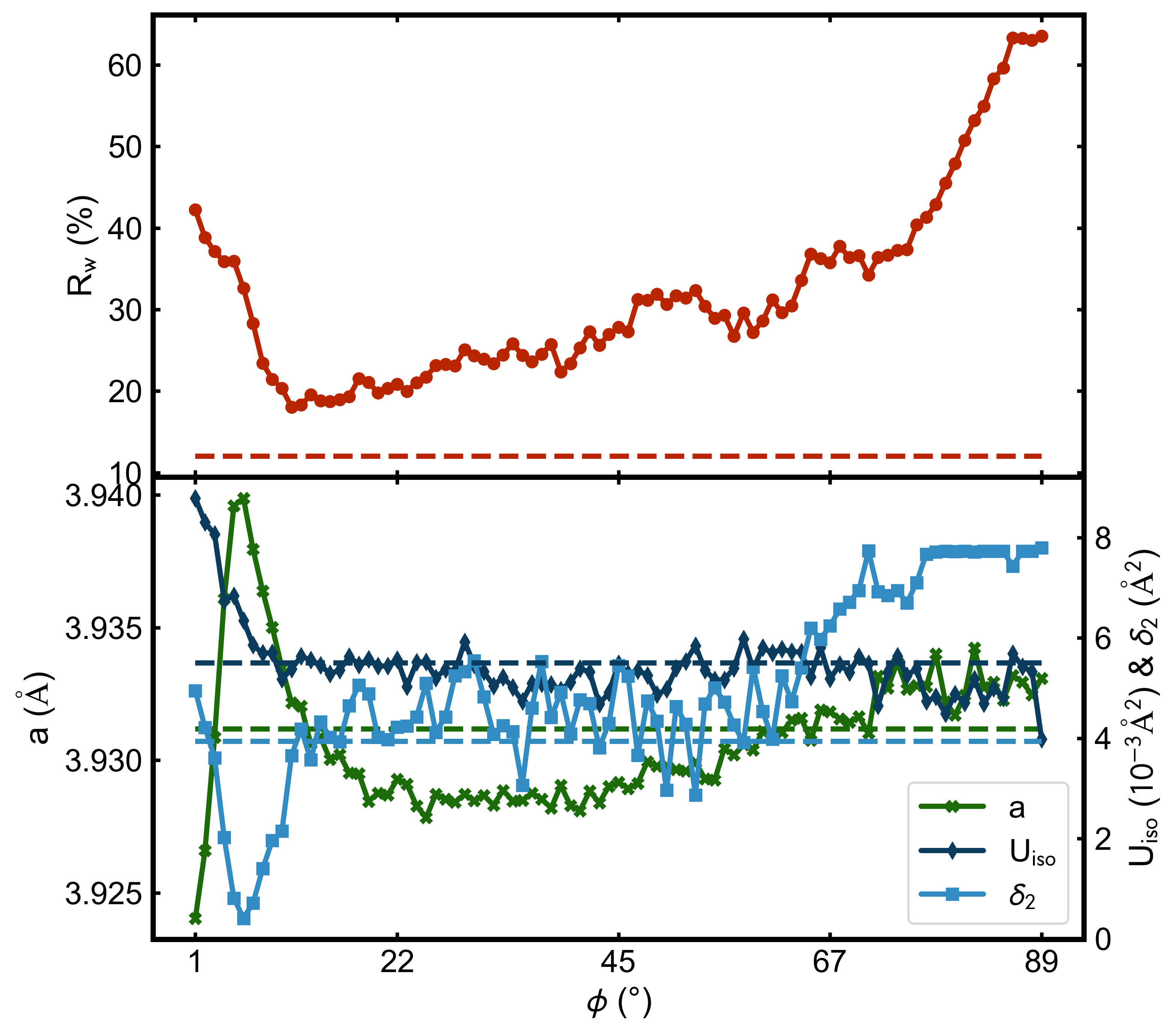}
		\caption{PDF refinement results of the 1D PDFs of the Platinum thin film at different tilt angles $\phi$.  The upper panel shows the goodness of the fit $R_{w}$, the lower panel shows the refined structural parameters the lattice parameter, a, the isotropic thermal displacement parameter $\mathrm{U_{iso}}$ and the coherent thermal motion parameter $\mathrm{\delta_{2}}$.  The dashed line shows the $R_{w}$ and the refined parameters of the averaged 1D PDF over the whole angular range $\phi$ = 1$^{\circ}$ to $\phi$ = 89$^{\circ}$.}
		\label{fig:pdf_ref_res}
	\end{figure}
	
	\begin{figure}
		\includegraphics[width=0.9\columnwidth]{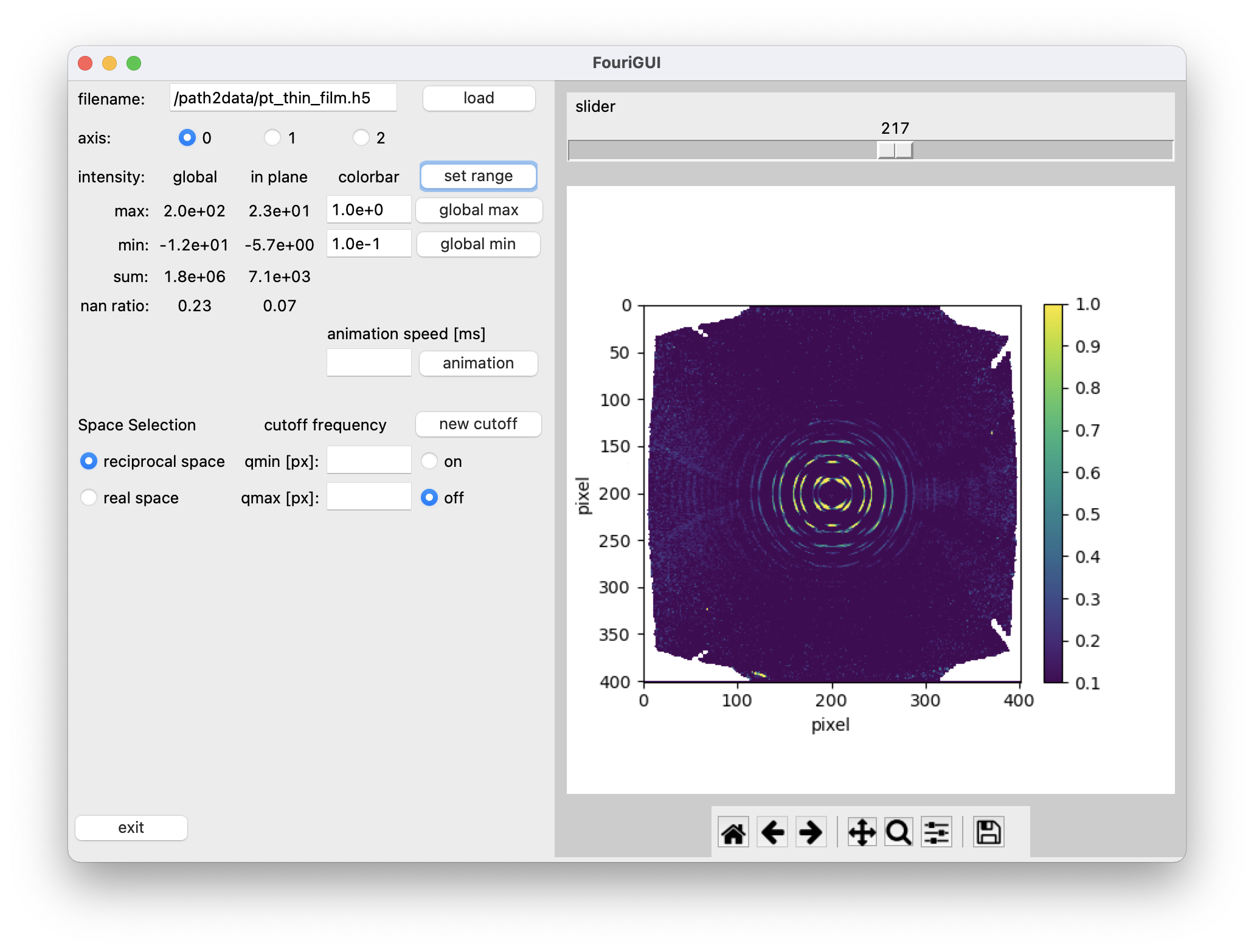}
		\caption{Image of FouriGUI in operation with a loaded total scattering structure function $S(\vec{Q})$ of the fiber-textured Pt thin film.  The data is loaded in h5py file format.  FouriGUI displays one plane of the loaded data perpendicular to a specified axis.  The slider on the top right enables the user to scroll through the data to visualize slices at different positions along that axis.  Fourigui shows global and in-plane live values of the maximum, minimum and summed intensity as well as a measure of the invalid pixels (containing non number values).  The Fourier transformation can be performed by changing to real space in the space selection.  One can further set cutoff frequencies and evaluate the effect directly in real and reciprocal space.  When the animation button is selected, FouriGUI scrolls through the images down the selected axis automatically.}
		\label{fig:screenshot_fourigui}
	\end{figure}

	

\end{document}